\documentclass[aps,prb,twocolumn,superscriptaddress,showpacs,floatfix]{revtex4-1}
\usepackage{amsmath}

\newif\ifprb
\prbtrue

\newcommand{\TN}{\ensuremath{T_{\it{N}}}}
\newcommand{\hw}{\ensuremath{\hbar \omega}}
\newcommand{\ibbs}[1]{\ensuremath{\boldsymbol{#1}}}
\newcommand{\imchiw}{\text{Im}\ensuremath{{\chi}(\hbar \omega)}}
\newcommand{\imchi}{\text{Im}\ensuremath{{\chi}}}
\newcommand{\etal}{{\it et al.}}
\newcommand{\gfemn}{$\gamma$-FeMn}
\newcommand{\cfemn}{Fe$_{0.7}$Mn$_{0.3}$}
\newcommand{\gcfemn}{$\gamma$-Fe$_{0.7}$Mn$_{0.3}$}
\newcommand{\mub}{\ensuremath{\mu_{\it{B}}}}
\newcommand{\invA}{{\AA}\ensuremath{^{-1}}}
\newcommand{\QAF}{\ensuremath{\boldsymbol{Q}_{\text{AF}}}}
\usepackage[dvipdfmx]{graphicx}

\usepackage[normalem]{ulem}
\usepackage[dvipdfmx]{color}

\begin{document}
\title{Damped spin-wave excitations in the itinerant antiferromagnet {\gcfemn}}

\author{S. Ibuka}
\email[]{ibuka.em@gmail.com}
\affiliation{Institute of Materials Structure Science, High Energy Accelerator Research Organization, Tsukuba 305-0801, Japan}
\affiliation{Materials \& Life Science Facility, J-PARC Center, Tokai, Ibaraki 319-1195, Japan}
\author{S. Itoh}
\author{T. Yokoo}
\affiliation{Institute of Materials Structure Science, High Energy Accelerator Research Organization, Tsukuba 305-0801, Japan}
\affiliation{Materials \& Life Science Facility, J-PARC Center, Tokai, Ibaraki 319-1195, Japan}
\affiliation{Department of Materials Structure Science, School of High Energy Science, Graduate University for Advanced Science, Tsukuba 305-0801, Japan}
\author{Y. Endoh}
\affiliation{Institute of Materials Structure Science, High Energy Accelerator Research Organization, Tsukuba 305-0801, Japan}
\affiliation{RIKEN Center for Emergent Matter Science, Wako 351-0198, Japan}
\affiliation{Department of Physics, Tohoku University, Aramaki Aoba, Sendai 980-8578, Japan}

\date{\today}

\begin{abstract}
The collective spin-wave excitations in the antiferromagnetic state of {\gcfemn} were investigated using the inelastic neutron scattering technique.
The spin excitations remain isotropic up to the high excitation energy, $\hw = 78$~meV. 
The excitations gradually become broad and damped above 40~meV. 
The damping parameter ${\gamma}$ reaches 110(16)~meV at $\hw = 78$~meV, which is much larger than that for other metallic compounds, e.g., 
CaFe$_2$As$_2$ (24~meV), La$_{2-2x}$Sr$_{1+2x}$Mn$_2$O$_7$ ($52-72$~meV), and Mn$_{90}$Cu$_{10}$ (88~meV).
In addition, the spin-wave dispersion shows a deviation from the relation $({\hw})^2 = c^2q^2 + {\Delta}^2$ above 40~meV.
The group velocity above this energy increases to 470(40)~meV{\AA}, which is higher than that at the low energies, $c = 226(5)$~meV{\AA}. 
These results could suggest that the spin-wave excitations merge with the continuum of the individual particle-hole excitations at 40~meV.
\end{abstract}


\maketitle

\section{Introduction}
There has been a renewed interest in magnetic excitations in itinerant antiferromagnets after the discovery of superconductivity in copper oxides~\cite{LeePA2006} and iron pnictides/chalcogenides~\cite{KamiharaY2008}. These superconductivities emerge near the antiferromagnetic ordered phases, and it is believed that spin fluctuations play a role in binding Cooper pairs.
Mechanism of the spin-wave excitations remains under debate in the iron pnictides/chalcogenides. The spin-wave excitations are highly in-plane anisotropic with a large damping at the high energy region ($> 100$~meV) of $A$Fe$_2$As$_2$ ($A =$ Ba, Sr or Ca)~\cite{HarrigerLW2011, EwingsRA2011, DialloSO2009}.
Various mechanisms were proposed to explain the in-plane anisotropy, which are based on local-moment $J_1-J_2$ models with electron nematic ordering~\cite{FangC2008} and orbital ordering~\cite{ChenCC2010}, itinerant models~\cite{KnolleJ2010, KaneshitaE2011, KovacicM2015}, and the combination of itinerant and localized characters~\cite{ParkH2011, YinZP2011}.
To have a profound understanding of the spin excitations in the superconductors and related compounds, we need some background knowledge about the spin excitations in prototypical itinerant antiferromagnets.
      
Metallic chromium, {\gfemn}, $\gamma$-Mn, and $\gamma$-Fe are prototypes of itinerant antiferromagnets. Chromium and {\gfemn} are gap-type antiferromagnets~\cite{AsanoS1971}, as well as iron pnictides~\cite{MazinII2008}. $\gamma$-Mn and $\gamma$-Fe are band-type antiferromagnets~\cite{AsanoS1971}. The gap-type antiferromagnetic states are induced by the nesting of the Fermi surfaces, while the band-type antiferromagnetic states are induced by the interband mixing due to the large spin-dependent perturbation~\cite{AsanoS1971}. 
In itinerant antiferromagnets, the maximum energy of the collective spin-wave excitations is relatively high compared that of Heisenberg magnets~\cite{KohgiM1980}, and thus the overall picture of the excitations has not yet been clarified~\cite{IshikawaY1978, EndohY2006}.
For example, the dynamic structure of chromium consists of incommensurate excitations below 20~meV~\cite{FincherCR1981} and commensurate excitations localized at the antiferromagnetic wave vector, which extends up to more than 550~meV~\cite{HeapRT1991, LowdenJR1995, HaydenSM2000}.
A recent theoretical study~\cite{SugimotoK2013} based on a multi-band Hubbard model predicted that the commensurate excitations merge with the continuum of the individual particle-hole excitations above 600~meV.
However, spin excitations above 600~meV have yet to be elucidated experimentally because inelastic neutron scattering experiments are difficult in this high-energy transfer region.

In particular, an experimental understanding of spin-wave damping due to the continuum of the particle-hole excitations is still lacking in itinerant antiferromagnets compared to that in itinerant ferromagnets~\cite{IshikawaY1978, EndohY2006}. 
To reveal how spin-wave excitations are damped at high energies in the gap-type itinerant antiferromagnets, we focused on {\gfemn}, whose spin-wave velocity is much smaller than of metallic chromium.
The Fe$_x$Mn$_{1-x}$ alloy ($0.3 < x < 0.85$) is crystallized in a face-centered cubic structure, the so-called $\gamma$-phase. The Fe and Mn ions are located randomly at the origin and face center positions.
{\gfemn} shows the antiferromagnetic transition in the temperature range $350 < T < 500$~K, depending on the composition $x$~\cite{KouvelJS1963, UmebayashiH1966, EndohY1971, IshikawaY1974}.
Asano {\etal}~\cite{AsanoS1971} revealed that the magnetic order is a spin-density-wave (SDW) order using a band calculation. The Fermi surfaces show good nesting with $\ibbs{Q} = (0, 0, 1)$ for $x = 0.4$. 
The magnetic structure is under debate experimentally and theoretically~\cite{KawarazakiS1990, KennedySJ1987, BisantiP1987, HiraiK1985, SakumaA2000, EkholmM2011}. The candidates are collinear single-SDW or non-collinear triple-SDW structures. Both Fe and Mn ions have magnetic moments. The average moment size is in the range $1\mub < \mu < 2 \mub$~\cite{KouvelJS1963, NathansR1964, IshikawaY1967, EndohY1971}, where {\mub} is the Bohr magneton.
Studies of the spin-wave dispersion up to 56~meV~\cite{TajimaK1976, EndohY1973} have reported that the spin-wave excitations follow the relation,
\begin{eqnarray}
\label{eq1}(\hw)^2 = c^2q^2 + \Delta^2,
\end{eqnarray}
where {\hw} is the excitation energy and $q$ is the distance from the antiferromagnetic zone center {\QAF}.
The spin-wave velocity is about $c = 280$~meV{\AA} and the energy gap about $\Delta = 9$~meV at room temperature (RT) for $x = 0.7$~\cite{TajimaK1976}.
Endoh {\etal}~\cite{EndohY1973} observed strong damping of the spin-wave excitations at $\hw = 56$~meV, possibly due to particle-hole excitations.
However, the dispersion and damping ratios are not known above this energy.
Therefore, in this study, we performed inelastic neutron scattering measurements on {\gfemn} to investigate the spin-wave excitations at high energies.
This investigation will enhance our understanding of spin excitations in itinerant antiferromagnets. In addition, the existence of the double-$Q$ magnetic states has been recently proposed in the tetragonal magnetic phase of hole-doped iron pnictides~\cite{WangX2015, SchererDD2016}. Spin-wave excitations and damping effects are keys to tell two possible double-$Q$ states~\cite{SchererDD2016}. This study could offer some insights into the double-$Q$ states of iron pnictides.

\section{Experimental details}
The composition $x = 0.7$ was selected, because the average moment size is the largest in the ${\gamma}$-phase.
A {\cfemn} crystal was synthesized by a Bridgman type induction furnace. The details have been described elsewhere~\cite{EndohY1971}. The mass of the grown single crystal was about 40~g. The antiferromagnetic ordering temperature for this composition is known to be $\TN = 435$~K with the average magnetic moment being 1.97\mub~\cite{IshikawaY1967}. The crystal orientation was determined using the x-ray Laue method (YXLON MG452, 450~kV/5~mA).
Inelastic neutron scattering experiments were performed on the High Resolution Chopper Spectrometer (HRC)~\cite{ItohS2011, HRC, ItohS2015} installed at the Materials and Life Science Facility, in the Japan Proton Accelerator Research Complex. DAVE/MSlice~\cite{dave} was used for analyzing the data. 
The initial neutron energies were set between $E_{\it i} =$ 33 and 372~meV. The energy resolution was determined using the incoherent elastic scattering of a solid vanadium, while the angular resolution was geometrically estimated.
The sample was mounted with a horizontal $hkk$ scattering plane, sealed in an aluminum can under a $^4$He gas atmosphere, and then set in a closed cycle $^4$He cryostat. Measurements were performed at $T = 14$~K. The lattice parameter was $a = 3.58$~{\AA} at that temperature. 
\begin{figure}
  \begin{center}
    \ifprb
    \hspace{-70px}
    \raisebox{-30px}[180px][-30px]{\includegraphics[width=0.98\hsize]{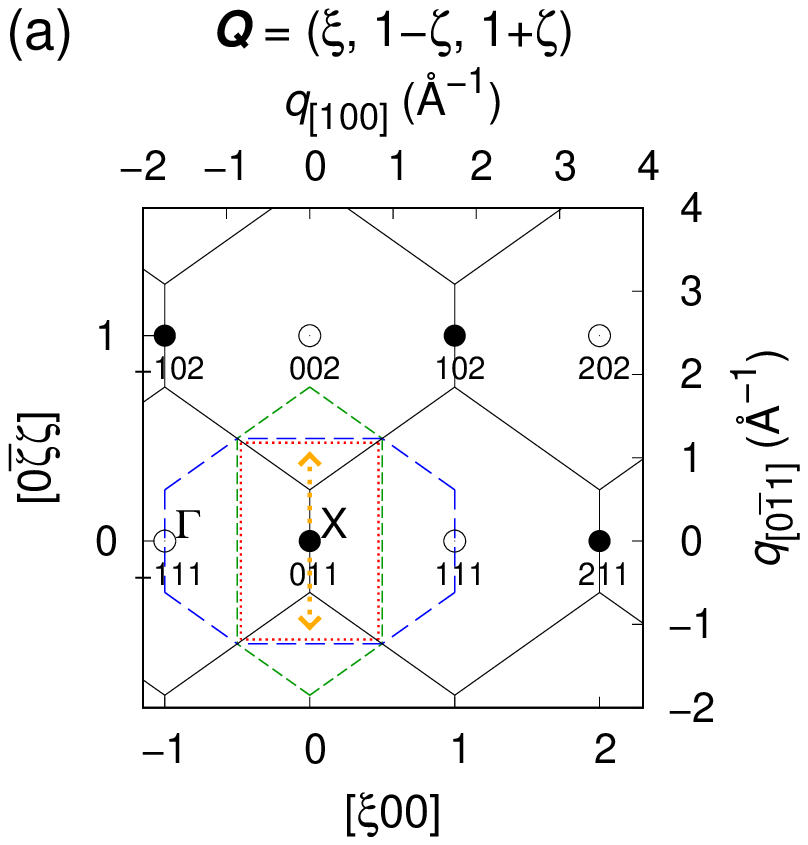}}\\
    \else
      \includegraphics[width=0.98\hsize]{fig1a.eps}\\
    \fi
  \includegraphics[width=0.7\hsize]{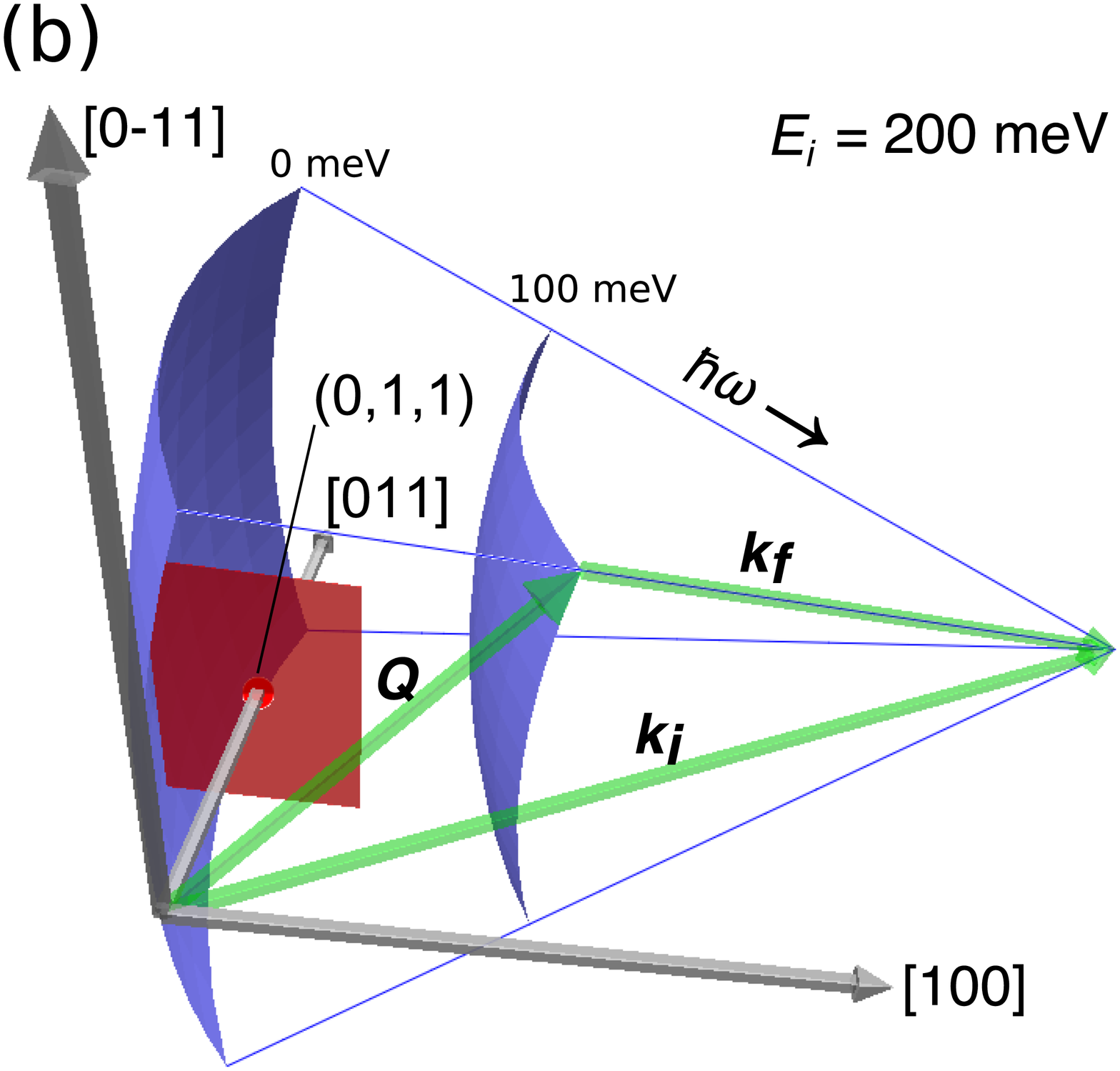}\\
  \caption{\label{fig1} (a) Brillouin zone perpendicular to [011] for {\gfemn}. The open and closed circles represent the nuclear and magnetic zone centers, respectively. The black solid lines represent the full zone. The green and blue dashed lines represent the reduced magnetic zones for the single-SDW structure. The red dotted lines represent the reduced magnetic zone for the triple-SDW structure. The orange dotted arrow shows the measurement direction for Figs.~\ref{fig2} and \ref{fig5}. (b) Typical measurement condition for the neutron scattering experiments described in the reciprocal space. The gray arrows represent the crystal directions for [100], [011], and [0$\bar{1}$1]. The red point represents the magnetic zone center ${\QAF} = (0, 1, 1)$. The red rectangular plate represents the reduced magnetic zone for the triple-SDW structure, which corresponds to the red dotted rectangle in (a). The green arrows represent the initial and final wave vectors of the neutron, $k_i$ and $k_f$, and the scattering vector {\ibbs{Q}} with $E_i = 200$~meV, respectively. 
The blue curved surfaces represent the measuring surfaces for the constant energy $\hw = 0$ and 100~meV.}
  \end{center}
\end{figure}
The Brillouin zone for {\gfemn} is shown in Fig.~\ref{fig1}(a). Antiferromagnetic zone centers are located at the symmetry point $X$. All the measurements were performed around the magnetic zone center $\QAF = (0, 1, 1)$. Figure~\ref{fig1}(b) shows a typical measurement condition for the neutron scattering measurements. By measuring the fixed crystal angle, the scattering intensity is obtained on the blue curved surfaces for each {\hw}. Thus, {\hw} of the measured intensity varies on the red rectangle zone. The scattering intensities for constant-\ibbs{Q} and -{\hw} can be obtained by several measurements with different crystal angles. In this study, measurements with rotating crystal angles were adopted to obtain the ${\hw} - q_{[0\bar{1}1]}$ map, which is shown in Fig.~\ref{fig2}. For the other measurements, fixed angle measurements were adopted due to the limited beam time.

\section{Results}
\begin{figure}
  \ifprb
  \hspace{-70px}
  \raisebox{-30px}[150px][-30px]{\includegraphics[width=0.98\hsize]{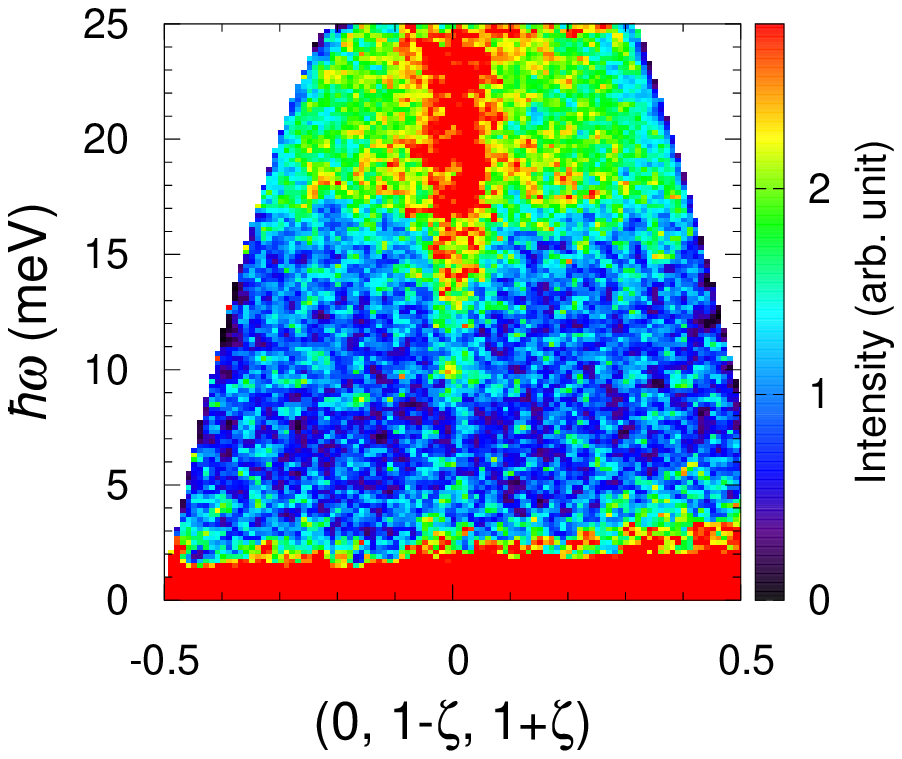}}\\
  \else
  \includegraphics[width=0.98\hsize]{fig2.eps}\\
  \fi
        \caption{\label{fig2} ${\hw} - q_{[0\bar{1}1]}$ map around {\QAF} at $T = 14$~K and $E_{\it i} = 33$~meV.}
\end{figure}
The ${\hw} - q_{[0\bar{1}1]}$ map around {\QAF} is shown in Fig.~\ref{fig2}. The measured \ibbs{Q} position corresponds to the orange dotted arrow in Fig.~\ref{fig1}(a). Spin-wave excitations with an energy gap of about 10~meV are found at {\QAF}. The uniform scattering above 15~meV is a background from aluminum polycrystals contained in the instruments and the sample can.

\begin{figure}
  \ifprb
  \hspace{-70px}
  \raisebox{-30px}[320px][-30px]{\includegraphics[width=0.98\hsize]{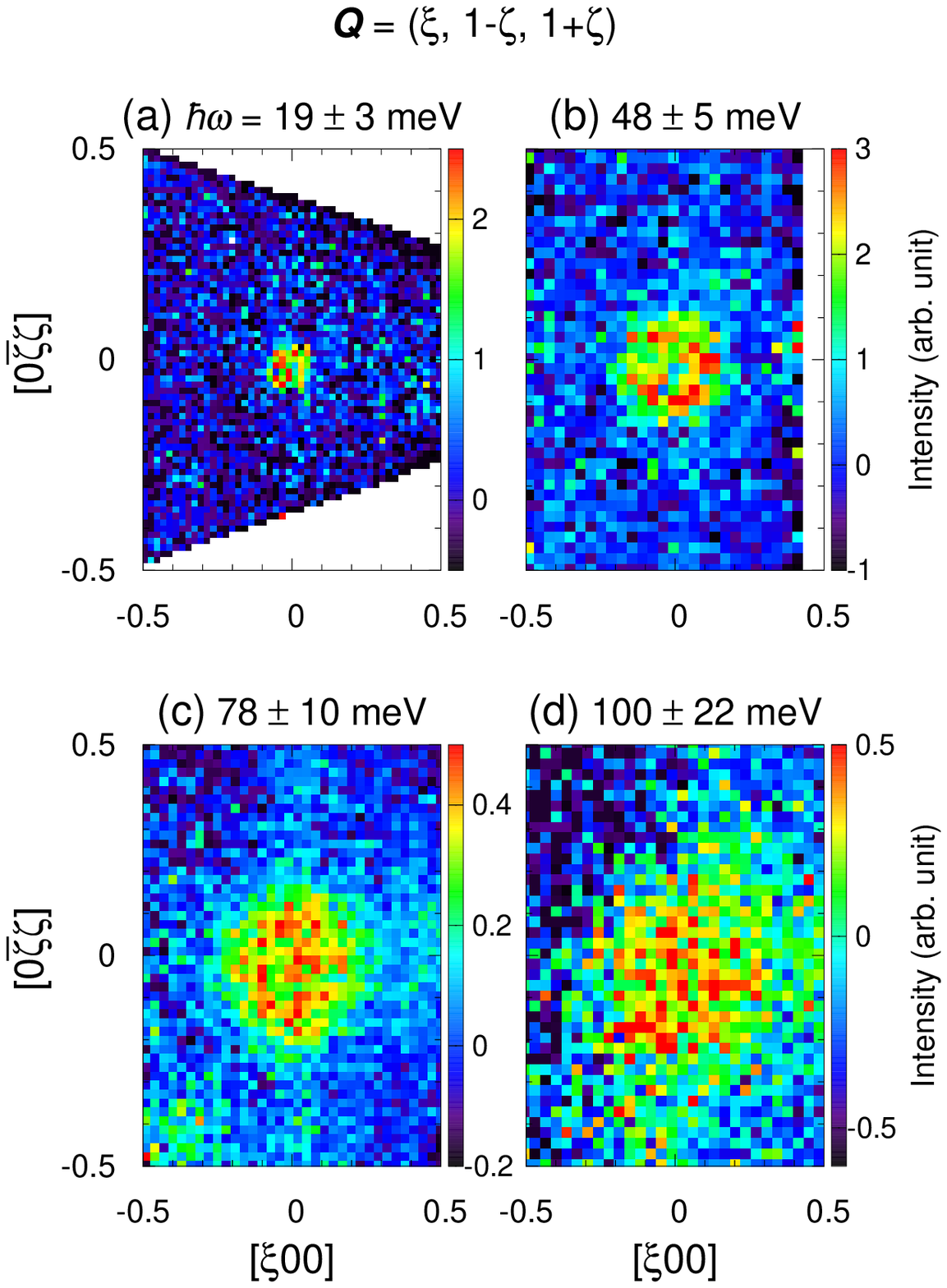}}\\
  \else
  \includegraphics[width=0.98\hsize]{fig3.eps}\\
  \fi
  \caption{\label{fig3} (a-d) $q_{[0\bar{1}1]} - q_{[100]}$ maps in {\cfemn} at $T = 14$~K, which show the reciprocal planes perpendicular to [011] centered at {\QAF}. 
    {\hw} = (a) $19 \pm 3$, (b) $48 \pm 5$, (c) $78 \pm 10$, and (d) $100 \pm 22$~meV. The errors of {\hw} denote the {\hw} range which the peaks cover. For example, for (a), energy resolution is $\pm 0.5$~meV. The average {\hw} for the left edge [-0.1 0 0], center, and right edge [0.1 0 0] of the peak is 16.5, 19, and 21.5~meV, respectively.
    $E_{\it i}$ = (a) 33, (b) 82, (c) 207, and (d) 372~meV. For clarity, background intensities were subtracted.}
\end{figure}
Figures~\ref{fig3}(a-d) show the $q_{[0\bar{1}1]} - q_{[100]}$ maps through {\QAF} with $\hw = 19$, 48, 78, and 100~meV, respectively. The measured \ibbs{Q} range corresponds to the red dotted rectangle in Fig.~\ref{fig1}(a). The edge of the panel is the zone boundary for the triple-SDW structure.
At (a) 19~meV, the magnetic excitations are localized at {\QAF} within $q < 0.1$~reciprocal lattice unit (r.l.u.)\ in all three directions [100], [110], and [111]. 
At (b) 48 and (c) 78~meV, the excitations open up in \ibbs{Q} space. Both excitations spread as circles, which indicate that the dispersion is isotropic. 
At (d) 100~meV, the excitations further spread compared to that at (c) 78~meV, but the spin-wave excitations do not reach the zone boundary even at this high energy. 
The excitations gradually become broad with increasing {\hw}.

\begin{figure}
  \ifprb
  \hspace{-70px}
  \raisebox{-30px}[170px][-30px]{\includegraphics[width=0.98\hsize]{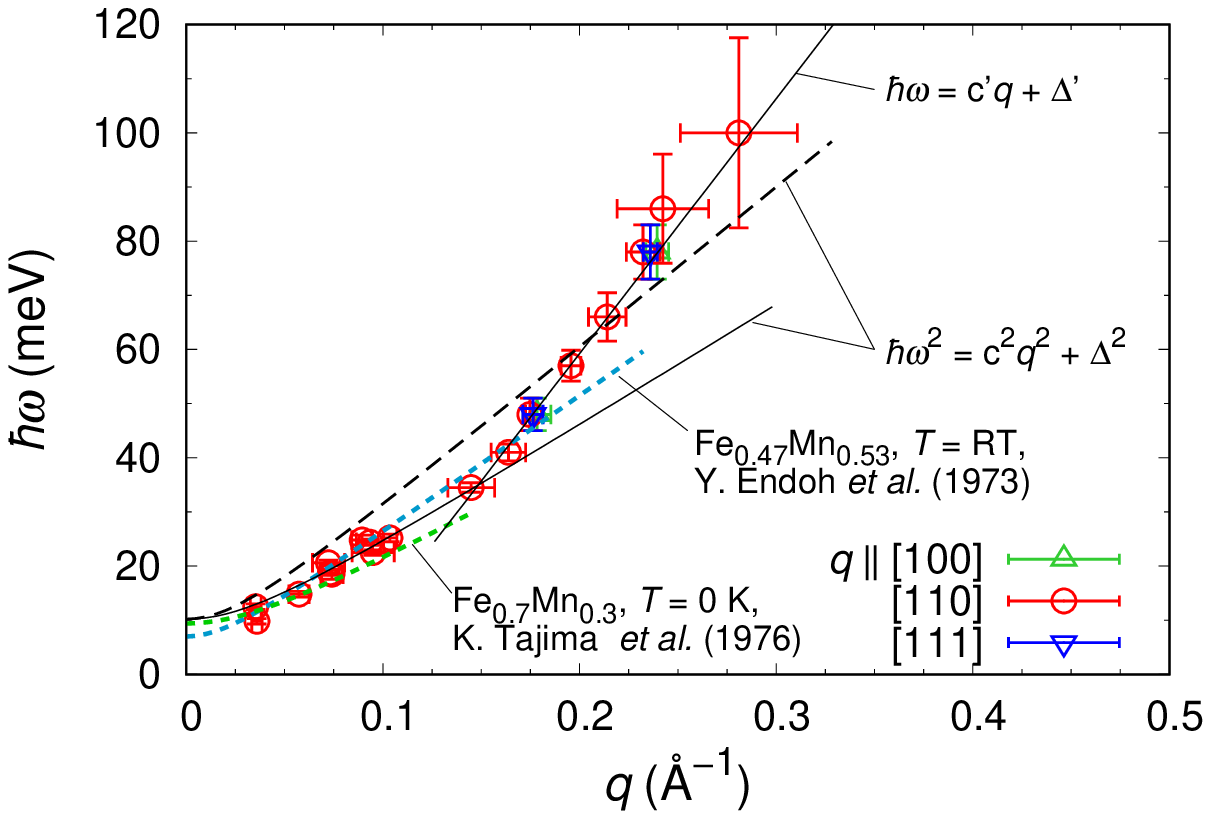}}
  \else
  \includegraphics[width=0.98\hsize]{fig4.eps}
  \fi
  \caption{\label{fig4} Dispersion relation of the spin-wave excitations in {\cfemn} at $T = 14$~K. The green triangles, red circles, and blue inverted triangles represent dispersion in [100], [110], and [111], respectively. The vertical error bars represent the energy resolution. The horizontal bars represent the fitting error. The black dashed and solid curves show fits with Eq.~\ref{eq1}. The black solid straight line shows a linear fit. The green and blue dotted curves show the dispersion relations previously reported for Fe$_{0.7}$Mn$_{0.3}$~\cite{TajimaK1976} at $T = 0$~K and Fe$_{0.47}$Mn$_{0.53}$~\cite{EndohY1973} at $T =$ RT, respectively. The nearest zone boundary is located at $q = 0.88$~\invA.}
\end{figure}
To obtain the dispersion relation quantitatively, the spin-wave excitations around {\QAF} were fitted to double Gaussians in [110]. 
In addition, to confirm the isotropy, the excitations were fitted in [100] and [111] in the same manner for $\hw = 48$ and 78~meV.
For the backgrounds, linear functions were used for [110], while cubic polynomial functions were used for [100] and [111] due to the volatile backgrounds in these directions.
Figure~\ref{fig4} shows the dispersion relation obtained by the fitting procedure.
At 48 and 78~meV, there is no difference in the peak positions between all the directions within the experimental error, which indicates that the dispersion is isotropic even at 78~meV.
This is consistent with the report~\cite{EndohY1973} that the dispersion is isotropic in [100] and [110] up to 35~meV.
Secondly, we tried to fit the dispersion relation with a function. The energy gap was fixed to $\Delta = 10.2$~meV in the following fitting procedure, which was obtained with the {\hw} dependence of the local spin susceptibility as described later.
The black dashed curve in Fig.~\ref{fig4} shows the fit with Eq.~\ref{eq1}, which indicates that the function is unsuited.
The dispersion could be fitted with Eq.~\ref{eq1} only below 40~meV. The black solid curve in Fig.~\ref{fig4} shows the fit below 40~meV. The spin-wave velocity is $c = 226(5)$~meV{\AA}.
Below this energy, the dispersion is reasonably consistent with those in the previous reports for Fe$_{0.7}$Mn$_{0.3}$ [$c = 195(30)$~meV{\AA}] at $T = 0$~K~\cite{TajimaK1976} and Fe$_{0.47}$Mn$_{0.53}$ ($245 - 265$~meV{\AA})~\cite{EndohY1973} at $T =$ RT, shown by the dotted curves in Fig.~\ref{fig4}.
Above 40~meV, the gradient of the dispersion relation is considerably larger than that expected from Eq.~\ref{eq1}. 
The linear fit with the function, ${\hw} = c'q + \Delta'$, above 40~meV yields the group velocity $c' = 470(40)$~meV{\AA}.

\begin{figure}
  \ifprb
  \hspace{-70px}
  \raisebox{-30px}[320px][-30px]{\includegraphics[width=0.98\hsize]{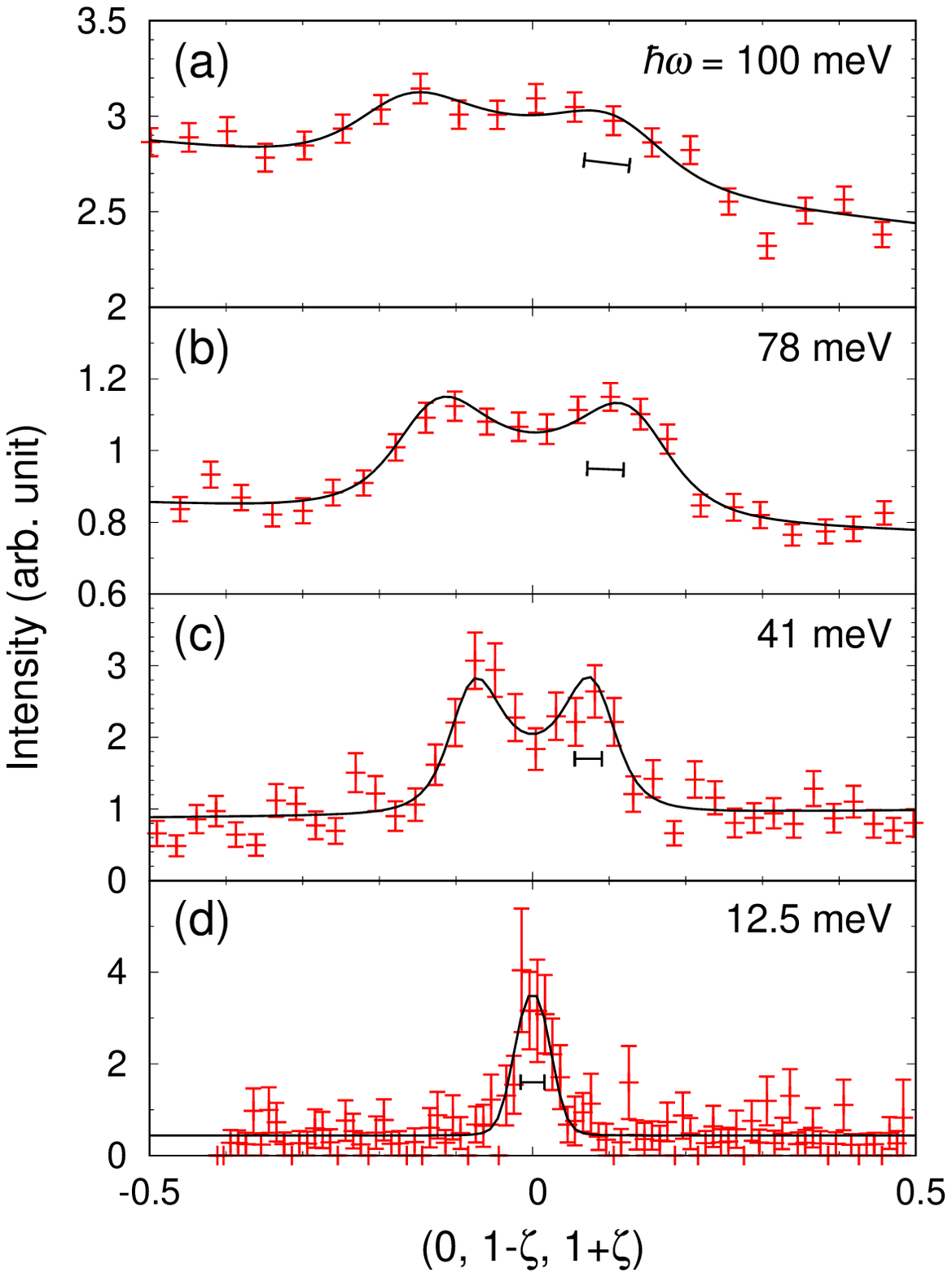}}
  \else
  \includegraphics[width=0.98\hsize]{fig5.eps}
  \fi
  \caption{\label{fig5} $q_{[0\bar{1}1]}$ cut at ${\hw} =$ (a) 100, (b) 78, (c) 41, and (d) 12.5~meV through {\QAF}. The measured $q$ position corresponds to the orange dotted line in Fig.~\ref{fig1}(a). The black bars represent the $q$ resolutions, respectively. The energy widths of the data points are (a) $\pm$18, (b) $\pm$7, (c) $\pm$1.6, and (d) $\pm$0.5~meV. $E_i =$ (a) 372, (b) 207, (c) 72, and (d) 33~meV.} The solid lines represent the fits.
\end{figure}
To obtain the lifetime of the excitations as a damping parameter ${\gamma}$,
the scattering intensity $I(q, \hw)$ along [0$\bar{1}$1] through {\QAF} was fitted to $S(q, \hw) k_f / k_i$ convoluted with the instrumental resolution adding a linear background in $q$.
In this analysis, the following equation based on the damped harmonic oscillator $\text{Im}\chi$ in $q_{[0\bar{1}1]}$ was assumed:
\begin{eqnarray}
  S(q, \hw) &\propto& \frac{\text{Im}{\chi}(q, \hw)}{1 - \exp \left[- \hw / (k_{\it B} T) \right] },\\
  \label{eq3}\text{Im}{\chi}(q, {\hw}) &\propto& \frac{{\hw}\gamma}{((\hw)^2 - (\hw(q))^2)^2 + (\hw)^2\gamma^2},
\end{eqnarray}
where $k_{\it B}$ is the Boltzmann constant. 
${\hw}(q)$ is the dispersion relation.
An appropriate ${\hw}(q)$ should be given to Eq.~\ref{eq3} before the fitting.
  We note that the data are not fitted well if ${\hw}(q)$ is assumed to be Eq.~\ref{eq1}.
  Here, ${\hw}(q)$ is assumed to be the dispersion relation extracted above.
The $q$ dependence of the magnetic form factor was ignored because of the small $q$ variation.
This analysis is equivalent to the earlier report~\cite{TajimaK1976}.
Figure~\ref{fig5} shows the fitting results for ${\hw} =$ 100, 78, 41, and 12.5~meV.
At all energies, the scattering intensities were fitted well with the above model function.
The observed peak widths are wider than the resolutions, and therefore, the peak broadening is an essential nature of the observed excitations.
\begin{figure}
  \ifprb
  \hspace{-70px}
  \raisebox{-30px}[320px][-30px]{\includegraphics[width=0.98\hsize]{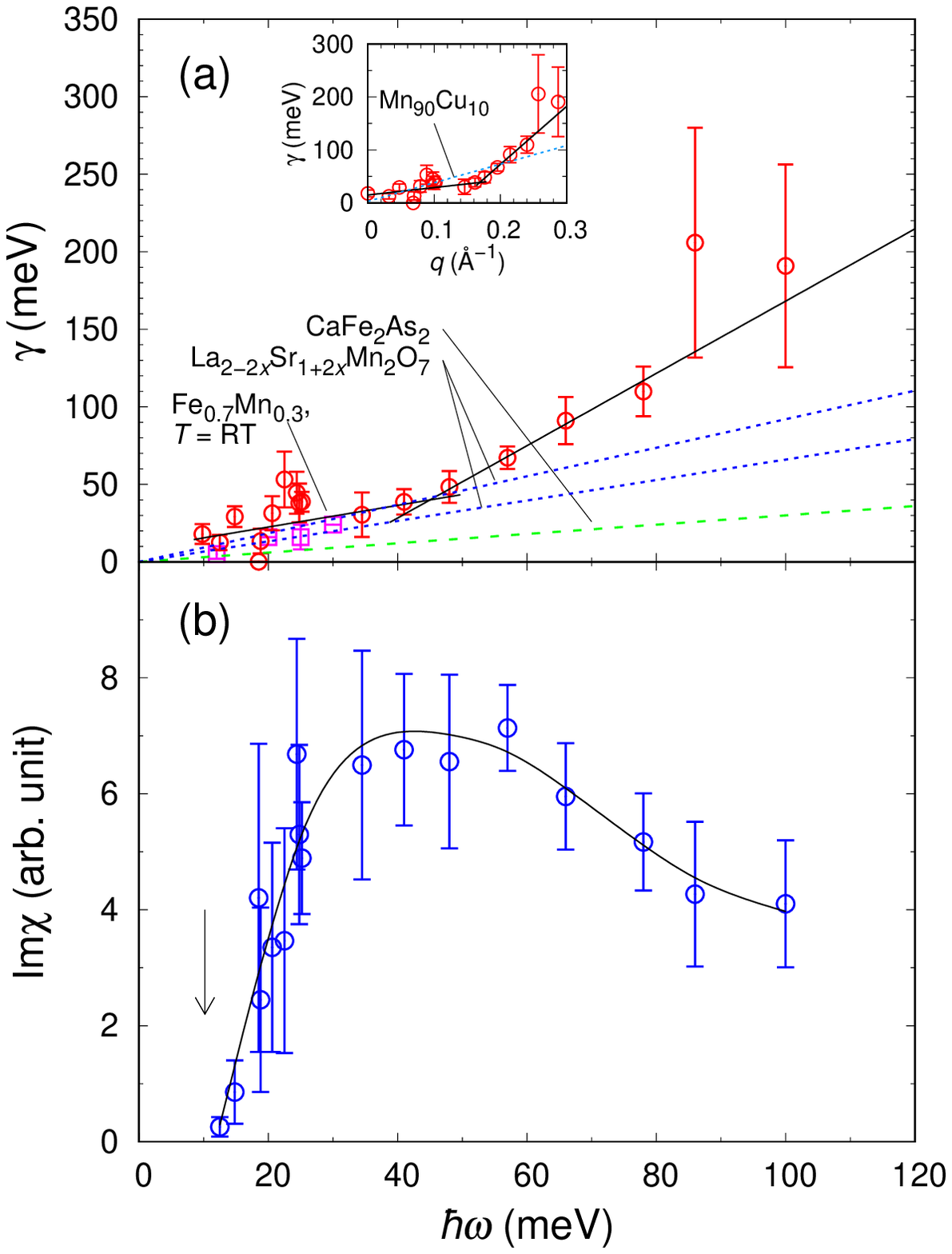}}
  \else
  \includegraphics[width=0.98\hsize]{fig6.eps}
  \fi
  \caption{\label{fig6} (a) {\hw} dependence of the damping parameter ${\gamma}$ for {\cfemn} at $T = 14$~K. The black solid lines are guides to the eyes.
    The purple squares show $\gamma$ previously reported for Fe$_{0.7}$Mn$_{0.3}$~\cite{TajimaK1976} at $T =$ RT.
    The green dashed and blue dotted lines show $\gamma$ for CaFe$_2$As$_2$~\cite{ZhaoJ2009} and La$_{2-2x}$Sr$_{1+2x}$Mn$_2$O$_7$~\cite{PerringTG2001}, respectively.
    $q$ dependence of ${\gamma}$ is shown in the inset. The sky-blue dotted line show $\gamma$ for Mn$_{90}$Cu$_{10}$~\cite{FernandezBacaJA1992}.
    (b) {\hw} dependence of the local spin susceptibility {\imchiw}. The black solid line is a guide to the eyes. The arrow represents the gap energy $\Delta$.}
\end{figure}
The obtained parameter ${\gamma}$ is shown in Fig.~\ref{fig6}(a). 
${\gamma}$ below 30~meV is in agreement with the previous report~\cite{TajimaK1976} at $T =$ RT.
${\gamma}$ shows steady increase above approximately 40~meV, and reaches 110(16)~meV at ${\hw} = 78$~meV.
${\gamma}$ below 40~meV is similar to that of the metallic compound, the bilayer colossal magnetoresistive manganites La$_{2-2x}$Sr$_{1+2x}$Mn$_2$O$_7$ (${\gamma}/{\hw} = 0.66 - 0.92$)~\cite{PerringTG2001}, but larger than that of another metallic compound, the parent compound of iron superconductor CaFe$_2$As$_2$ (0.3)~\cite{ZhaoJ2009}.
${\gamma}$ above 40~meV is much larger than those of both compounds.
The inset of Fig.~\ref{fig6}(a) shows $q$ dependence of $\gamma$. $\gamma$ increases steadily above $q = 0.16$~{\invA}. In Mn$_{90}$Cu$_{10}$, which is a dilute alloy of the band-type itinerant antiferromagnet $\gamma$-Mn, $\gamma$ was found to be linear in $q$~\cite{FernandezBacaJA1992} as shown by the sky-blue dotted line. $\gamma$ is of the form $\gamma(q) = \gamma_0 + \gamma_1 q$, where $\gamma_0 = 3$~meV and $\gamma_1 = 351$~meV{\AA} below 0.8~{\invA}~\cite{FernandezBacaJA1992, FernandezBacaJA1993}. An increase in the rate of change of $\gamma$ was not observed in these three compounds.
Note that, in these reports, ${\gamma}$ is determined in the same manner as this study under the assumption of a damped simple harmonic oscillator.

Next, the local spin susceptibility {\imchiw}, which is approximately proportional to the integrated intensity, was derived using the following equations,
\begin{eqnarray}
  \label{eqint}{\imchiw} &=& \int \text{Im}{\chi}(\ibbs{q}, \hw) \text{d}\ibbs{q},\\
  &\simeq& \int_{\text{BZ}} \frac{k_{\it i}}{k_{\it f}} \left[ I\left(\ibbs{q}, \hw \right) - B \right]\nonumber\\
  && \hspace{2em} \times \left\{ 1 - \exp \left[- \hw / \left(k_{\it B} T\right) \right] \right\} d\ibbs{q},
\end{eqnarray}
where the integration was performed within a magnetic Brillouin zone. The energy dependence on \ibbs{q}, which came from the measurements with the fixed crystal angle, was ignored because the gradient of the dispersion relation is large. 
$B$ is the background function, which was determined using the intensity away from {\QAF}. For example, intensity around $\ibbs{Q} = (0, 0.5, 1.5)$ was used for the background at $\ibbs{Q} = (0, 1, 1)$.
Figure~\ref{fig6}(b) shows {\imchiw}.
The energy gap was estimated to be ${\Delta}$(14~K) $= 10.2(7)$~meV by linear extrapolation of {\imchi} below 15~meV. The result is consistent with the value previously reported~\cite{TajimaK1976}, $\Delta$(0~K) $= 9.4$~meV, which was derived from the temperature dependence of ${\Delta}$.
{\imchi} reaches the maximum value around 40~meV and clearly decreases above 60~meV.

\section{Discussion}
The high group velocity and isotropic dispersion of the spin-wave excitations in {\cfemn} are common to the commensurate spin excitations in chromium and its dilute alloys.
The spin-wave velocity in dilute alloys of chromium is more than 1000~meV{\AA}~\cite{AlsNielsenJ1971, SinhaSK1977} if the commensurate excitations are assumed to be spin-wave excitations. In pure chromium, \ibbs{Q} width of the commensurate mode remains constant from 350 to 550~meV~\cite{LowdenJR1995}. The spin-wave velocities are at least an order of magnitude larger than that in {\gfemn}.
This suggests that the Fermi velocities of the nested bands are smaller in {\gfemn} than those in chromium and its dilute alloys~\cite{FeddersPA1966}. This is consistent with the large thermal effective electron mass in {\gfemn} observed in the electronic specific heat measurements~\cite{GuptaRP1964, HashimotoT1967}.
In itinerant magnets, dispersions are isotropic in many cases.
However, an anisotropic dispersion may be observed near the zone boundaries if the spin wave remains well defined, as calculated for Fe, Co, and Ni~\cite{HalilovSV1998}.
The isotropic result will not determine the spin structure of {\gfemn}. With a localized spin model~\cite{JensenJ1981}, an isotropic dispersion is expected at low $q$ for the triple-SDW spin structure. However, Tajima {\etal} reported~\cite{TajimaK1976} that the scattering profiles are inconsistent with those of the calculation. Theoretical works based on itinerant electron models will be needed. Note that Sato and Maki studied~\cite{SatoH1976} the spin fluctuations with a two-band model near the transition temperature, but the analysis was performed under the condition of spherical Fermi surfaces. 

The energy gap could originate by the spin-orbit interaction. The spin-orbit scenario expects that SDW has an orbital character. The existence of the spin-orbit coupling was suggested by Ishikawa {\etal}~\cite{IshikawaY1973} They found the anisotropy of the critical scattering in {\gfemn}. 
The origin of the large energy gap might have a relationship with Weyl fermions. It is suggested that the Berry phase induces the orbital ferromagnetism and anomalous Hall effect for the triple-SDW spin structure in {\gfemn} when the crystal structure is distorted~\cite{ShindouR2001, FangZ2003}, and the effect of the Berry phase is observable with inelastic neutron scattering~\cite{ItohS2016}. Although either ferromagnetism or crystal distortion have not yet been observed in {\gfemn}, the large energy gap and the anisotropic critical scattering imply the existence of the Berry phase. To elucidate of the origin of the large energy gap in {\gfemn}, further investigations are required.

It is really strange that the group velocity of the spin wave suddenly increases at 40~meV and that the spin-wave dispersion deviates from Eq.~\ref{eq1}.
This indicates the existence of other excitations that interact with the spin wave, because the group velocity decreases with energy if the spin wave does not interact.
This result strongly suggests that the spin wave merges with the continuum of the individual particle-hole excitations at 40~meV.
When a spin wave enters the continuum, an increase in the group velocity is expected for a ferromagnetic electron gas~\cite{IshikawaY1978}. On the other hand, a magnon-electron interaction will not cause a sudden change.
The relatively large $\gamma$ is in agreement with the origin of the damping. 
{\imchi} gradually decreases above 60~meV and $\gamma$ steadily increases above 40~meV, which indicates that the spin wave gradually damps with increasing {\hw} in the continuum.
A similar {\hw} dependence of {\imchi} and $\gamma$ was reported at the boundary of the particle-hole excitations in the weak itinerant ferromagnets MnSi~\cite{IshikawaY1977}.
Note that because the neutron scattering cross section of the particle-hole excitations is quite small, it is difficult to measure the excitations directly.
Of course, other scenarios for explaining the excitations cannot be ruled out.
The site randomness may be another possibility. The inhomogeneously distributed moments caused by the randomly located Fe and Mn ions~\cite{KimballC1963, IshikawaY1967} could disturb propagation of the spin wave.
At least, it is unique that the spin-wave excitations show the linear $q$ dependent dispersion and broad energy width above 40~meV. A future theoretical study would be very interesting.
  
If the unique behavior of the spin-wave excitations originates in the interaction with the particle-hole excitations, the similar behavior will be expected also in other itinerant antiferromangets.
In chromium and its dilute alloys, the large spin-wave velocity makes it difficult to detect the deviation of the dispersion relation from Eq.~\ref{eq1}. Thus, to reveal the influence of particle-hole excitations on the spin wave in Cr, {\imchi} is important.
The experimental data of {\imchi} for the commensurate excitations are controversial. {\imchi} reported by Fukuda {\etal}~\cite{FukudaT1996} decreases above 60~meV, the energy scale of which is similar to that of this study. On the other hand, Booth {\etal}~\cite{BoothJG1988} reported that {\imchi} is constant between 48 and 120~meV.
A precise study on {\imchi} in chromium will be required to investigate the spin-wave damping due to particle-hole excitations.
For the band-type itinerant antiferromagnets, a spin-wave calculation predicted that the damping is linear in $q$ at small $q$ due to the decay of the spin-wave excitations into particle-hole pairs~\cite{GillanMJ1973}. This theory shows semiqualitative agreement with the experimental results in Mn$_{90}$Cu$_{10}$~\cite{FernandezBacaJA1992, FernandezBacaJA1993}. A similar weak damping roughly linear in $q$ was observed in {\gfemn} below 40~meV. This damping will be due to the weak decay into the particle-hole pairs below the continuum and due to the site randomness. 

\section{Summary}
Inelastic neutron scattering measurements were performed to investigate high-energy magnetic excitations in {\cfemn}. The spin-wave velocity suddenly increases at $\hw = 40$~meV. The local spin susceptibility gradually decreases above 60~meV and the damping parameter increases above 40~meV, which demonstrates that the spin wave gradually damps with increasing the excitation energy.
This study could provide an important example of the spin wave damping due to the particle-hole excitations in itinerant antiferromagnets.

\begin{acknowledgments}
The authors are indebted to T. Masuda for providing the x-ray Laue instruments at the Institute for Solid State Physics, the University of Tokyo. 
The neutron scattering experiment at HRC was approved by the Neutron Scattering Program Advisory Committee of the Institute of Materials Structure Science, High Energy Accelerator Research Organization (Nos.\ 2014S01, 2015S01).
\end{acknowledgments}

\bibliography{femn}

\end{document}